\documentclass[12pt]{revtex4}
\usepackage{epsfig,url}
\newcommand\eps{\varepsilon}

\newcommand\Real{\mathop{\mathrm{Re}}}
\newcommand\Imag{\mathop{\mathrm{Im}}}
\newcommand\diag{\mathop{\mathrm{diag}}}

\newcommand\bk{\bar\kappa}
\begin{document}
\title{Emittance growth of kicked and mismatched beams due to amplitude-dependent tune shift}
\author{E. Waagaard\footnote{presently at CERN.}, V. Ziemann}
\affiliation{Department of Physics and Astronomy, Uppsala University, S-75120 Uppsala, Sweden}
\date{\today}
\begin{abstract}\noindent
  We derive evolution equations for the first and second moments of an initially
  mismatched, coupled, and displaced arbitrary Gaussian phase-space distribution
  under the influence of decoherence due to amplitude-dependent tune shift. Moreover,
  we find expressions for the asymptotic values of the beam matrix and the emittance
  and use them to evaluate error tolerances for injection.
\end{abstract}
\maketitle
%
%
\section{Introduction}
The emittance of a beam, injected into a ring, crucially depends on the initial position
and angle of the injected beam as well as on the Twiss parameters of the injection line
being equal to those of the ring. Once the beam is circulating in the ring, the particles
perform betatron oscillations around the equilibrium orbit in the ring. Any spread of
betatron frequencies, either due to chromaticity and a finite momentum spread, or due
to amplitude-dependent tune shift, causes the distribution of particles to distort and
evolve into one with a larger emittance. This process is often referred to as decoherence.
This decoherence of kicked beams due to amplitude-dependent tune shift was previously
analyzed in~\cite{Meller,SYLEE,Sargsyan} where, however, only the decoherence of the
centroid was evaluated. Moreover, in~\cite{MINTY} the evolution of the kicked beam matrix
is calculated and the key results are summarized in~\cite{HANDBOOK}. Here we extend
the analysis by considering the turn-by-turn evolution of the first and second moments
of a beam that initially is both displaced and mismatched. We then follow the evolution
of its first moments, which are often referred to as centroids, as well as its beam
matrix and emittance, as the beam decoheres. 
\par
In order to prepare the stage for our calculations, we assume that the optics in the
ring is uncoupled. We therefore introduce the phase shift per turn $\phi_x$ in the
horizontal plane due to normal betatron phase advance $\mu_x=2\pi Q_x$ and to
amplitude-dependent tuneshift, given by
\begin{equation}\label{eq:adps}
  \phi_x = \mu_x + \kappa_{xx} \left(x_1^2+x_2^2\right) +  \kappa_{xy} \left(x_3^2+x_4^2\right) 
  =\mu_x+\vec x^{\top}\bar\kappa_x\vec x\ ,
\end{equation}
where $\vec x^{\top}$ is the transpose of $\vec x$ and
$\bar\kappa_x=\diag(\kappa_{xx},\kappa_{xx},\kappa_{xy},\kappa_{xy})$. Here $\kappa_{xx}$
parameterizes the amplitude dependence in the horizontal plane and $\kappa_{xy}$ its dependence
on the amplitude in the vertical plane, also called the cross-anharmonicity~\cite{VERDIER}.
Here $2J_x=x_1^2+x_2^2=\gamma_x  x^2+2\alpha_x x x' +\beta_x x^{\prime 2}$ with
$\gamma_x=(1+\alpha_x^2)/\beta_x$ is twice the Courant-Snyder invariant $J_x$ of the
linear motion in the horizontal plane and $2J_y=x_3^2+x_4^2=\gamma_y  y^2+2\alpha_y y y'
+\beta_y y^{\prime 2}$ in the vertical plane.
We use variables $x_1,\dots x_4$ in normalized phase space, collectively denoted by
$\vec x=(x_1,x_2,x_3,x_4)^{\top}.$ They are related to the position $x$ and angle $x'$ by
\begin{equation}\label{eq:calA}
  \left(\begin{array}{c} x_1 \\ x_2 \end{array}\right)
  = {\cal A}_x
  \left(\begin{array}{c} x \\ x' \end{array}\right)
  \qquad\mathrm{with}\qquad
  {\cal A}_x=\left(\begin{array}{cc} 1/\sqrt{\beta_x} & 0 \\
                     \alpha_x/\sqrt{\beta_x} & \sqrt{\beta_x}\end{array}\right)\ ,
\end{equation}
where $\alpha_x$ and $\beta_x$ are the Twiss parameters in the horizontal plane of the
ring at the point of injection. In most of this report we henceforth focus on the horizontal plane.
The corresponding equations for the coordinates in the other plane $x_3$ and $x_4$ the
subscript $x$ is exchanged with subscript $y$. Note also that after $n$ revolutions in the ring,
the phase shift is $n\phi_x$. In passing we point out that it is straightforward to generalize
Equation~\ref{eq:adps} to six dimensions by adding a term $\kappa_{xs}(x_5^2+x_6^2)$,
extending the definition of $\bar\kappa_x$ to a $6\times 6$ matrix that includes $\kappa_{xs}$
on the two lowest entries on the diagonal and interpreting $\vec x$ as the corresponding
six-dimensional phase space vector. In this report, however, we focus on two and four
dimensions.
\par
We always assume that the initial beam distribution is a multivariate Gaussian. For convenience
we define it as the $d$-dimensional distribution
\begin{equation}\label{eq:gauss}
  \psi_d\left(\vec x;\vec X,\sigma\right)=\frac{1}{(2\pi)^{d/2}\sqrt{\det\sigma}}
  \exp\left[ -\frac{1}{2}\sum_{j,k=1}^d \sigma^{-1}_{jk}(x_j-X_j)(x_k-X_k)\right]\ ,
\end{equation}
where $d$ can be 2 or 4, depending on the phase space we consider. Moreover,
$X_j$ with $j=1,\dots,d$ are the components of the vector $\vec X$ with the initial
centroid positions. The $d\times d$ matrix $\sigma$ is the beam matrix describing
the widths and orientations of the Gaussian. Note that in coordinates of normalized
phase space, the beam matrix $\sigma$ of a matched beam in all planes is proportional
to the unit matrix. For a matched beam, the proportionality constant
in each $2\times 2$ block on the diagonal is the emittance of the injected beam
$\eps_{0}$ in the respective plane. Throughout this report we normalize positions
and beam sizes by $\sqrt{\eps_0}$, such that all numerical values are given in units
of the corresponding rms values of the beam size or the angular divergence. For
example, the physical position $x$ is related to $x_1$ through $x_1=x/\sqrt{\beta}$
and normalized by $\sqrt{\eps_0}$ to $x/\sqrt{\eps_0\beta}$.
\par
In the following sections, we first follow the centroid of this Gaussian as it
decoheres, where we assume that $\sigma$ is an arbitrary beam matrix, not
necessarily matched to the ring into which we assume the beam is injected.
In Section~\ref{sec:ampl}, we show that our general
result reproduces the results from~\cite{Meller} for a matched injected beam.
In the following sections, we calculate the turn-by-turn evolution of the second
moments in general, before considering a matched beam and an arbitrary beam
matrix in one transverse plane. In Section~\ref{sec:coup} we consider injection
of a transversely coupled beam matrix. In all cases we derive expressions
for the asymptotic beam matrix and then use them to determine error tolerances.
In separate sections, we discuss the asymptotic emittance growth due to a
mismatched dispersion and indicate how to include decoherence due to chromaticity
into our framework before summarizing our results in the conclusions.
\section{Centroid}
\label{sec:centroid}
We now calculate the betatron motion with phase advance $\mu_x$ of the centroid of
a Gaussian and denote the centroid position in the horizontal plane after $n$ turns by
$\hat X_1$ and $\hat X_2$, which leads us to
\begin{equation}\label{eq:centa}
 \hat X_1+i\hat X_2
  =e^{-in\mu_x}\left\langle e^{-in\vec x^{\top}\bar\kappa_x\vec x}(x_1+ix_2)\right\rangle
\end{equation}
where the angle brackets denote averaging over the initial Gaussian distribution from
Equation~\ref{eq:gauss}. We point out that damping can be taken into account by adding
a factor $e^{-n/N_d}$ (with damping time given in number of turns $N_d$) to the right-hand
side of Equation~\ref{eq:centa}. But in this report we do not pursue this further.
Since we will encounter similar integrals to those appearing in Equation~\ref{eq:centa}
along the way, we introduce the notation
\begin{equation}\label{eq:INP}
I[n,p] = \left\langle e^{-in\vec x^{\top}\bar\kappa_x\vec x}p(\vec x)\right\rangle\ ,
\end{equation}
where $p(\vec x)$ is a multi-variate polynomial in the phase-space coordinates
$x_1,\dots x_d.$ In Equation~\ref{eq:centa}, for example, we have $p(\vec x)=x_1+ix_2$.
Moreover, Equation~\ref{eq:centa} can also be expressed as $\hat X_1+i X_2=
e^{-in\mu_x}I[n,x_1+ix_2]$. 
\par
In the next step, we evaluate $I[n,p]$ by explicitly writing it as a Gaussian integral
\begin{equation}\label{eq:Inp}
  I[n,p] = \frac{1}{(2\pi)^{d/2}\sqrt{\det\sigma}} \int d^dx\,
  e^{-\frac{1}{2}\sum_{j,k=1}^d \sigma^{-1}_{jk}(x_j-X_j)(x_k-X_k)}
  e^{-in\vec x^{\top}\bar\kappa_x\vec x}
\end{equation}
where, for brevity, we suppress the limits of the integrals, which always extend from
$-\infty$ to $\infty$. We simplify the integrand by expressing $x_1^2$ as
\begin{equation}
x_1^2=(x_1-X_1)^2+2 X_1 x_1-X_1^2 = (x_1-X_1)^2 + 2X_1(x_1-X_1)+X_1^2
\end{equation}
and likewise for $x_2^2,\dots,x_d^2$. Inserting in Equation~\ref{eq:Inp} and
combining terms, we arrive at
\begin{equation}\label{eq:Inp2}
  I[n,p]=\frac{e^{-in\vec X^{\top}\bar\kappa_x\vec X}}{(2\pi)^{d/2}\sqrt{\det\sigma}} \int d^dx\, 
            e^{-\frac{1}{2}\sum_{j,k=1}^d (\sigma^{-1}_{jk}+2in(\bar\kappa_x)_{kj})(x_j-X_j)(x_k-X_k)}
            e^{-2in\vec X^{\top}\bar\kappa_x(\vec x-\vec X)} p(\vec x)\ .
\end{equation}
We now introduce the abbreviations
\begin{equation}\label{eq:AB}
  A_{jk}=\sigma_{jk}^{-1}+2in(\bar\kappa_x)_{jk}
  \qquad\mathrm{and}\qquad
  B_j=2n\sum_{k=1}^d(\bar\kappa_x)_{jk} X_k\ .
\end{equation}
The substitution $\vec y= \vec x - \vec X$ then allows us to write Equation~\ref{eq:Inp2} as
\begin{equation}
  I[n,p]=\frac{e^{-in\vec X^{\top}\bar\kappa_x\vec X}}{(2\pi)^{d/2}\sqrt{\det\sigma}} \int d^dy\, 
  e^{-\frac{1}{2}\sum_{j,k=1}^d A_{jk}y_jy_k-i\sum_{j=1}^d B_j y_j} p(\vec y+\vec X)\ .
\end{equation}
In the final step, we find a substitution that helps us to remove the term
that is linear in $y_j$ in the exponent. We therefore introduce a further substitution
$z_j=y_j+h_j$ and find $h_j$ that removes that term. We insert this substitution into
the exponent and obtain
\begin{eqnarray}\label{eq:exp}
  &&-\frac{1}{2}\sum_{j,k=1}^d A_{jk}(z_j-h_j)(z_k-h_k)-i\sum_{j=1}^d B_j(z_j-h_j)\\
  &&\qquad=-\frac{1}{2}\sum_{j,k=1}^dA_{jk}z_jz_k+i\sum_{j=1}^dB_jh_j-\frac{1}{2}A_{jk}h_jh_k
     +\sum_{j=1}^d\left[\frac{1}{2}\sum_{k=1}^d 2A_{jk}h_k-iB_j\right]z_j\nonumber
\end{eqnarray}
which implies that
\begin{equation}
h_k=i\sum_{j=1}^dA^{-1}_{kj}B_j
\end{equation}
makes the square bracket zero and thus removes the linear term. After substituting
$h_k$ into the right-hand side of Equation~\ref{eq:exp}, the exponent assumes the form
\begin{equation}
-\frac{1}{2} \vec B^{\top} A^{-1}\vec B - \frac{1}{2}\sum_{j,k=1}^dA_{jk}z_jz_k\ .
\end{equation}
For $I[n,p]$ we find
\begin{eqnarray}\label{eq:IInp}
  I[n,p]&=&\frac{e^{-in\vec X^{\top}\bar\kappa_x\vec X-\frac{1}{2}\vec B^{\top} A^{-1}\vec B }}{(2\pi)^{d/2}\sqrt{\det\sigma}}
            \int d^dz\, e^{-\frac{1}{2}\sum_{j,k=1}^d A_{jk}z_jz_k}p(\vec x)\\
        &=& \frac{e^{-in\vec X^{\top}\bar\kappa_x\vec X-2n^2\vec X^{\top}\bar\kappa_x(\mathbf{1}+2in\sigma\bar\kappa_x)^{-1}\sigma\bar\kappa_x\vec X}}
            {(2\pi)^{d/2}\sqrt{\det\sigma}}
            \int d^dz\, e^{-\frac{1}{2}\sum_{j,k=1}^d A_{jk}z_jz_k}p(\vec x)
            \nonumber
\end{eqnarray}
with
\begin{equation}
  \vec x = \vec z+\vec X-iA^{-1}\vec B =\vec z + \vec Y
  \qquad\mathrm{and}\qquad
  \vec Y=\left(\mathbf{1}-2in A^{-1}\bar\kappa_x\right)\vec X\ .
\end{equation}
Moreover, we use the definitions of $A$ and $\vec B$ from Equation~\ref{eq:AB} to obtain
\begin{equation}\label{eq:YY}
  A^{-1}\vec B =2n(\mathbf{1}+2in\sigma\bar\kappa_x)^{-1}\sigma\bar\kappa_x\vec X
  \quad\mathrm{and}\quad 
  \vec Y=\left(\mathbf{1}+2in\sigma\bar\kappa_x\right)^{-1}\vec X\ .
\end{equation}
The integrals are evaluated with the help of the identities~\cite{ZJ}
\begin{eqnarray}\label{eq:integrals}
  \int d^dz\, e^{-\frac{1}{2}\sum_{j,k=1}^d A_{jk}z_jz_k} &=& \frac{(2\pi)^{d/2}}{\sqrt{\det A}}\nonumber\\
  \int d^dz\, e^{-\frac{1}{2}\sum_{j,k=1}^d A_{jk}z_jz_k} z_m &=& 0\\
  \int d^dz\, e^{-\frac{1}{2}\sum_{j,k=1}^d A_{jk}z_jz_k} z_mz_n &=&  \frac{(2\pi)^{d/2}}{\sqrt{\det A}}  A^{-1}_{mn}
  \nonumber
\end{eqnarray}
which follow from the well-known identities for normalizing a Gaussian distribution, and how the
first and second moments are given in terms of the covariance matrix. In particular, the
centroid positions after $n$ turns $\hat X_1+i\hat X_2$, identified by a caret, turn out to be
\begin{eqnarray}\label{eq:evoX}
  \hat X_1+i\hat X_2
  &=&e^{-in\mu_x} \frac{e^{-in\vec X^{\top}\bar\kappa_x\vec X-\frac{1}{2}\vec B^{\top} A^{-1}\vec B }}{(2\pi)^{d/2}\sqrt{\det\sigma}}
      \frac{(2\pi)^{d/2}}{\sqrt{\det A}} \left(Y_1+iY_2\right)\\
  &=& \frac{e^{-in\mu_x-in\vec X^{\top}\bar\kappa_x\vec X-2n^2\vec X^{\top}\bar\kappa_x(\mathbf{1}+2in\sigma\bar\kappa_x)^{-1}\sigma\bar\kappa_x\vec X}}
      {\sqrt{\det(\mathbf{1}+2in\sigma\bar\kappa_x)}}\left(Y_1+iY_2\right)\ ,\nonumber
\end{eqnarray}
where $\vec Y$ is defined in Equation~\ref{eq:YY}.
We point out that the result in Equation~\ref{eq:evoX} is valid for dimensions $d=2$ or $4$ and
for arbitrary beam matrices $\sigma$, including matched beams. In order to compare with the
results from~\cite{Meller}, we consider such a matched beam for $d=2$ in the following section.
\section{Amplitude dependence}
\label{sec:ampl}
In order to obtain some intuition, we compare our calculation with~\cite{Meller} and set
$d=2$ and $\kappa_{xx}=\kappa$ before calculating the evolution of the oscillation
amplitude of the centroid $a_n$ with the number of turns $n$
\begin{equation}
  a_n=\sqrt{\vert\vec{\hat X}\vert^2}=\sqrt{\hat X_1^2+\hat X_2^2} 
     =\sqrt{\left(\hat X_1+i\hat X_2\right)\left(\hat X_1-i\hat X_2\right)}
\end{equation}
for a matched beam with the $2\times 2$ beam matrix
\begin{equation}
  \sigma=\eps_0\mathbf{1}\ .
\end{equation}
To do so, we take the squared modulus of Equation~\ref{eq:evoX} and consider one term at a time.
First, we consider $\vec Y$ and calculate $\vert\vec Y\vert^2$ from Equation~\ref{eq:YY},
which leads to
\begin{equation}
  \vert\vec Y\vert^2
  =\left(\frac{1}{1+2in\kappa\eps_0}\right)\left(\frac{1}{1-2in\kappa\eps_0}\right)\vert\vec X\vert^2
  =\frac{1}{1+4n^2\kappa^2\eps_0^2}\vert\vec X\vert^2\ .
\end{equation}
Second, we consider the root in the denominator of Equation~\ref{eq:evoX}, which simplifies to
\begin{equation}
\sqrt{\det(\mathbf{1}+2in\kappa\sigma)}=\sqrt{\det((1+2in\kappa\eps_0)\mathbf{1})}=1+2in\kappa\eps_0
\end{equation}
which has squared modulus $1+4n^2\kappa^2\eps_0^2$ that consequently also appears in the denominator. 
Finally, the third term in the exponent of Equation~\ref{eq:evoX} simplifies to
\begin{equation}
  \vec X^{\top}(\mathbf{1}+2in\kappa\sigma)^{-1}\sigma\vec X
  = \vec X^{\top}\frac{\eps_0}{1+2in\kappa\eps_0}\vec X
  = \frac{\eps_0\vert\vec X\vert^2}{1+4n^2\kappa^2\eps_0^2}(1-2in\kappa\eps_0)\ .
\end{equation}
Since the imaginary part in the exponent has unit modulus, only the real part appears 
in the modulus of the whole expression. Inserting the three contributions into 
Equation~\ref{eq:evoX} results in
\begin{equation}\label{eq:Xhat2}
  \vert\vec {\hat X}\vert^2 = \frac{\vert\vec X\vert^2}{(1+4n^2\kappa^2\eps_0^2)^2}
        \exp\left[-\frac{4n^2\kappa^2\eps_0\vert\vec X\vert^2}{1+4n^2\kappa^2\eps_0^2}\right]\ .
\end{equation}
Expressing this equation in terms of the amplitude $a_n$ with the initial amplitude
$a_0=\sqrt{|\vec X|^2}$, we find
\begin{equation}\label{eq:amp}
a_n=\frac{a_0}{1+4n^2\kappa^2\eps_0^2}
\exp\left[-\frac{a_0^2}{2\eps_0}\ \frac{4n^2\kappa^2\eps_0^2}{1+4n^2\kappa^2\eps_0^2}\right]\ ,
\end{equation}
which agrees with the result for the amplitude decoherence from~\cite{Meller} provided
we identify $\theta=2n\kappa\eps_0$ and $\eps_0=1$.
\par
\begin{figure}[tb]
\begin{center}
\includegraphics[width=0.7\textwidth]{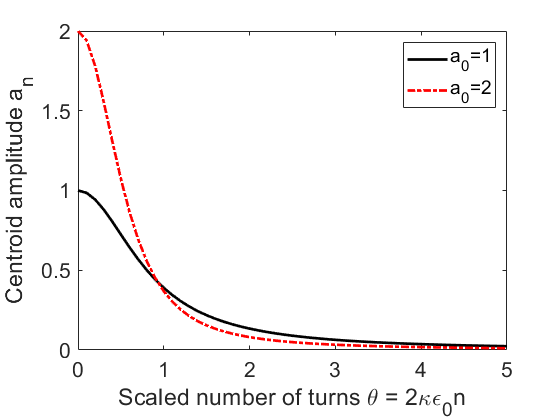}
\end{center}
\caption{\label{fig:amp}Amplitude of the beam centroid (in units of $\sqrt{\eps_0}$) versus the turn
  number $n$, parameterized as $\theta=2\kappa\eps_0 n$ with parameters $\eps_0=1$ and $\kappa=0.1$
  and for two values $a_0=1$ and $a_0=2$ of the initial displacement.}
\end{figure}
In Figure~\ref{fig:amp}, we use Equation~\ref{eq:amp} to show the dependence of the 
amplitude $a_n$ on the number of turns for starting amplitudes $a_0=\eps_0$ and 
$a_0=2\eps_0.$  We observe that  the initial reduction of the amplitude follows a 
Gaussian behavior, whereas for large $n$ the exponential approaches $e^{-a_0^2/2\eps_0}$ and the 
turn-evolution is governed by the factor $1+4n^2\kappa^2\eps_0^2$ in the denominator.
The transition between the two regimes, already discussed in~\cite{Meller},
appears around $\theta\approx1$ when $n\approx1/2\kappa\eps_0.$
A larger starting amplitude $a_0=2$ (red dashed curve) leads to a 
faster initial reduction of the amplitude to values below those for $a_0=1$ (black solid 
curve). Note that the curves cross near the transition at $\theta\approx 1.$ 
\par
In the next section, we turn to the evolution of the beam matrix and the emittance.
\section{Beam matrix and emittance}
In this section, we consider the general case with $d$ dimensions.
The beam size after $n$ turns is related to the second moments of the distribution
after $n$ turns, again identified by a caret. One of the moments $\langle \hat x_1^2\rangle$
is given by 
\begin{equation}
  \langle \hat x_1^2\rangle = \left\langle\left(x_1\cos n\phi_x+x_2\sin n\phi_x\right)^2\right\rangle\ .
\end{equation}
The angle brackets denote averaging over the initial distribution from
Equation~\ref{eq:gauss} in $d$ dimensions and $\phi_x$ is defined in Equation~\ref{eq:adps}.
All other moments, such as $\langle\hat x_1\hat x_2\rangle$ and $\langle \hat x_2^2\rangle$
are given by similar equations. We now express the trigonometric functions by their
exponential representation and arrive at
\begin{eqnarray}\label{eq:x1sq}
  \langle \hat x_1^2\rangle
  &=& \frac{1}{4}\left\langle\left(2+e^{2in\phi_x}+e^{-2in\phi_x}\right) x_1^2
      -2i\left(e^{2in\phi_x}-e^{-2in\phi_x}\right) x_1x_2\right.\nonumber\\
&& \qquad  \left.    +\left(2-e^{2in\phi_x}-e^{-2in\phi_x}\right) x_2^2\right\rangle\ .
\end{eqnarray}
At this point we note that only expressions of the type $e^{-im\phi}$ with $m=0,$ $2n,$
and $-2n$ appear. We therefore introduce
\begin{equation}\label{eq:Jmp}
  J[m,p;\mu_x,\bar\kappa_x]=\langle e^{-im\phi_x} p(\vec x)\rangle
  =\langle e^{-im\mu_x-im\vec x^{\top}\bar\kappa_x\vec x} p(\vec x)\rangle
\end{equation}
where $p(\vec x)$ is one of $x_1^2$, $x_1x_2$, $x_2^2$. For brevity, we omit the arguments
after the semicolon if they are unambiguous and just write $J[m,p]$. In the next step, we use
Equation~\ref{eq:Jmp} to rewrite $\langle \hat x_1^2\rangle$ in Equation~\ref{eq:x1sq},
which leads us to
\begin{eqnarray}\label{eq:x1J}
  \langle \hat x_1^2\rangle
  &=& \frac{1}{4}\left( 2J[0,x_1^2]+J[-2n,x_1^2]+J[2n,x_1^2] \right.\nonumber\\
  &&\quad    -2i J[-2n,x_1x_2] + 2i J[2n,x_1x_2]\nonumber\\
  && \quad\left.  +2J[0,x_2^2]-J[-2n,x_2^2]-J[2n,x_2^2]\right)\\
  &=&\frac{1}{2}\left(J[0,x_1^2]+\Real(J[-2n,x_1^2])\right) +\Imag(J[-2n,x_1x_2])\nonumber\\
  &&\quad +\frac{1}{2}\left(J[0,x_2^2]-\Real(J[-2n,x_2^2])\right)\ ,\nonumber
\end{eqnarray}
where we use
\begin{equation}
  J[-m,p]+J[m,p]=2 \Real(J[-m,p])
  \quad\mathrm{and}\quad
  J[-m,p]-J[m,p]=2i\Imag(J[-m,p])\ .
\end{equation}
The corresponding expressions for $\langle \hat x_1\hat x_2\rangle$, $\langle \hat x_2^2\rangle$,
and  $\langle \hat x_1\hat x_3\rangle$ can be found in Appendix~\ref{sec:appA}. 
\par
In order to evaluate $J[m,p]$, we note that it is closely related to $I[m,p]$ from
Equation~\ref{eq:Inp}, which allows us to express $J[m,p]$ as
\begin{equation}\label{eq:JI}
  J[m,p]= e^{-im\mu_x}\langle e^{-im\vec x^{\top}\bar\kappa_x \vec x}p(\vec x)\rangle = e^{-im\mu_x}I[m,p]\ .
\end{equation}
This leaves us the task to evaluate  $I[m,p]$ for $p=x_rx_s$ where $r$ and $s$ assume
values between $1$ and $d$. Expressing $x_r$ through $x_r=z_r+Y_r$ and inserting this in
Equation~\ref{eq:IInp}, we obtain
\begin{eqnarray}
  I[m,x_rx_s]
  &=&  \frac{e^{\psi(m)}}{(2\pi)^{d/2}\sqrt{\det\sigma}}
      \int d^dz\, e^{-\frac{1}{2}\sum_{j,k=1}^d A_{jk}z_jz_k}(z_r+Y_r)(z_s+Y_s)\\
  &=& \frac{e^{\psi(m)}}{(2\pi)^{d/2}\sqrt{\det\sigma}}
      \int d^dz\, e^{-\frac{1}{2}\sum_{j,k=1}^d A_{jk}z_jz_k}(z_rz_s+z_rY_s+z_sY_r+Y_rY_s)
      \nonumber
\end{eqnarray}
with the abbreviation 
\begin{equation}\label{eq:psim}
  \psi(m)=-im\vec X^{\top}\bar\kappa_x\vec X-2m^2\vec X^{\top}\bar\kappa_x(\mathbf{1}+2im\sigma\bar\kappa_x)^{-1}\sigma\bar\kappa_x\vec X\ .
\end{equation}
The four terms inside the integral are evaluated by using the expressions from
Equation~\ref{eq:integrals} and this leads to
\begin{eqnarray}
  I[m,x_rx_s]
  &=& \frac{e^{\psi(m)}}{(2\pi)^{d/2}\sqrt{\det\sigma}} \frac{(2\pi)^{d/2}}{\sqrt{\det A}} \left( A_{rs}^{-1}+Y_rY_s \right)\nonumber\\
  &=& \frac{e^{\psi(m)}}{\sqrt{\det(\mathbf{1}+2im\sigma\bar\kappa_x)}}\left( A_{rs}^{-1}+Y_rY_s \right)
\end{eqnarray}
and for $J[m,x_rx_s]$ we obtain with Equation~\ref{eq:JI}
\begin{equation}\label{eq:Jmrs}
J[m,x_rx_s]=\frac{e^{-im\mu_x+\psi(m)}}{\sqrt{\det(\mathbf{1}+2im\sigma\bar\kappa_x)}}\left( A_{rs}^{-1}+Y_rY_s \right)\ .
\end{equation}
with $A^{-1}=(\mathbf{1}+2im\sigma\bar\kappa_x)^{-1}\sigma$ and $\vec Y=(\mathbf{1}+2im\sigma\bar\kappa_x)^{-1}\vec X$.
\par
The matrix elements of the beam matrix after $n$ turns $\hat\sigma_{rs}$ are related to the
second moments $\langle \hat x_r\hat  x_s\rangle$ via
\begin{equation}\label{eq:sigmahat}
  \hat\sigma_{rs}=\langle(\hat x_r-\hat X_r)(\hat x_s-\hat X_s)\rangle
  = \langle \hat x_r\hat x_s\rangle - \hat X_r\hat X_s\ ,
\end{equation}
which requires us also to subtract $\hat X_r \hat X_s$ from the second
moments for which we resort to Equation~\ref{eq:evoX} to calculate $\hat X_1$ and  $\hat X_2$.
Both the second moments and the centroids must be calculated for the same number of turns $n$.
These equations are valid for any mismatched and transversely coupled beam that additionally
is injected off-axis with $\vec X\ne 0$.
\section{Emittance growth for a matched beam}
Just as we did for the amplitude decoherence, we now consider $d=2$, set $\kappa_{xx}=\kappa$,
and evaluate the turn-by-turn evolution of
the second moments and the emittance for a matched beam with $\sigma=\eps_0\mathbf{1}$,
analogous to the analysis from~\cite{MINTY}. We start our analysis by evaluating
the terms that enter $J[m,x_rx_s]$. The first is
\begin{equation}
\left(\mathbf{1}+2im\kappa\sigma\right)^{-1}=\frac{1}{1+2im\kappa\eps_0}\mathbf{1}
\end{equation}
which leads us to 
\begin{equation}
\vec Y = \left(\mathbf{1}+2im\kappa\sigma\right)^{-1}\vec X = \frac{1}{1+2im\kappa\eps_0}\vec X
\end{equation}
and 
\begin{equation}
A^{-1}= \left(\mathbf{1}+2im\kappa\sigma\right)^{-1}\sigma= \frac{\eps_0}{1+2im\kappa\eps_0}\mathbf{1}\ .
\end{equation}
The root in the denominator of Equation~\ref{eq:Jmrs} simplifies to
\begin{equation}
\sqrt{\det(\mathbf{1}+2im\kappa\sigma)}=1+2im\kappa\eps_0
\end{equation}
and $\psi(m)$ from Equation~\ref{eq:psim} becomes
\begin{eqnarray}
\psi(m)&=&-im\kappa(X_1^2+X_2^2)-2m^2\kappa^2\vec X^{\top} \frac{\eps_0}{1+2im\kappa\eps_0}\vec X\nonumber\\
       &=&-im\kappa\vert \vec X\vert^2-2m^2\kappa^2\frac{\eps_0}{1+2im\kappa\eps_0}\vert \vec X\vert^2\\
       &=&-\frac{im\kappa}{1+2im\kappa\eps_0}\vert \vec X\vert^2\ .  \nonumber
\end{eqnarray}
Inserting these expressions into Equation~\ref{eq:Jmrs}, we find
\begin{eqnarray}\label{eq:Jmrsr}
  J[m,x_rx_s]
  &=&\frac{e^{-im\mu_x-\frac{im\kappa}{1+2im\kappa\eps_0}\vert \vec X\vert^2}}{1+2im\kappa\eps_0}
  \left(\frac{\eps_0}{1+2im\kappa\eps_0}\delta_{rs}+\frac{X_rX_s}{(1+2im\kappa\eps_0)^2}\right)\nonumber\\
  &=&\frac{e^{-im\mu_x-\frac{im\kappa}{1+2im\kappa\eps_0}\vert \vec X\vert^2}}{(1+2im\kappa\eps_0)^2}
      \left(\eps_0\delta_{rs} +\frac{X_rX_s}{1+2im\kappa\eps_0}\right)
\end{eqnarray}
that we use to calculate the second moments from Equations~\ref{eq:x1J} and~\ref{eq:xxJ}.
\par
For the beam matrix we also need the centroid motion that we previously analyzed
in Section~\ref{sec:centroid} and for a matched beam in Section~\ref{sec:ampl}.
Adapting Equation~\ref{eq:evoX} to $\sigma=\eps_0\mathbf{1}$, we arrive at
\begin{equation}\label{eq:centr}
  \hat X_1+i\hat X_2
  = \frac{e^{-in\mu_x-\frac{in\kappa}{1+2in\kappa\eps_0}\vert \vec X\vert^2}}{(1+2in\kappa\eps_0)^2} (X_1+iX_2)
\end{equation}
whose modulus again leads to Equation~\ref{eq:Xhat2}. We emphasize that here $n$ is the
number of turns and not a general parameter such as $m$ in Equation~\ref{eq:Jmrsr}.
\par
\begin{figure}[tb]
\begin{center}
\includegraphics[width=0.97\textwidth]{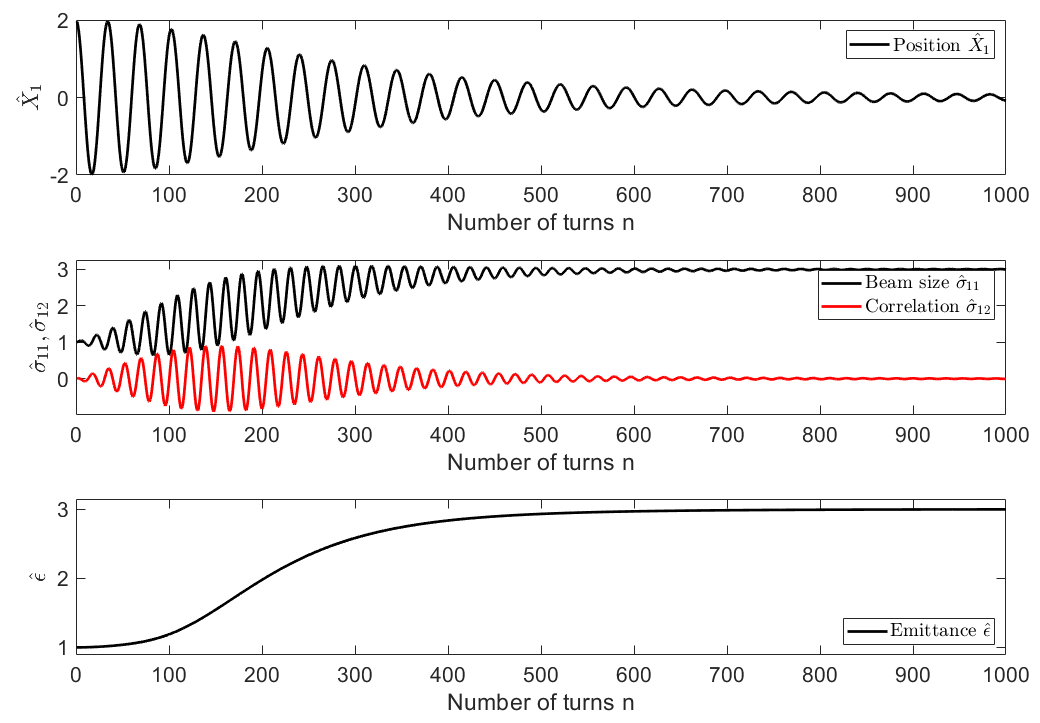}
\end{center}
\caption{\label{fig:emit}The centroid $X_1$ (top), the beam matrix elements $\sigma_{11}$
  and $\sigma_{12}$ (middle), and the emittance (bottom) as a function of the turn number $n$
  for a matched beam that is injected with initial offset $X_1=2.$ The parameters used
  are $\mu/2\pi=0.028$ and $\kappa\eps_0=0.001$. The vertical axes are normalized
  to appropriate powers of $\eps_0$.}
\end{figure}
From the second moments from Equation~\ref{eq:xxJ}, together with $J[m,x_rx_s]$ from
Equation~\ref{eq:Jmrsr} and the centroid from Equation~\ref{eq:centr}, we prepared a
MATLAB~\cite{MATLAB} script, available from~\cite{GITHUB}, to follow the centroids $\vec{\hat X}$,
the beam matrix $\hat\sigma$ from Equation~\ref{eq:sigmahat}, and the emittance
$\hat \eps=\sqrt{\det\hat\sigma}$ for a number of turns. Figure~\ref{fig:emit} shows
$\hat X_1$ (top), $\hat\sigma_{11}$ and $\hat\sigma_{12}$ (middle) and the emittance
$\hat\eps$ (bottom) as a function of $n$. The parameters in this simulation, chosen
to illustrate the dynamics, are $\mu_x/2\pi=0.028$, $\kappa=0.001$, and $\eps_0=1$.
Initially the beam is offset
by $X_1=2$ and the top plot shows oscillations that initially follow a Gaussian behavior
before later decaying at a much slower rate, as discussed in Section~\ref{sec:ampl}.
At the same time the beam size $\hat\sigma_{11}$ oscillates at twice the frequency of the
centroid and increases towards a higher level. Intermittently the correlation
$\hat\sigma_{12}$ increases, which is due to distortions of the initially matched beam
while it decoheres. Towards the end of the simulation, $\hat\sigma_{12}$ decreases to
zero, because the beam decoheres and reaches its equilibrium configuration. The
bottom plot shows the emittance $\hat\eps$, which has tripled compared
to the initially injected beam.
\par
The equilibrium value that is reached after the decoherence has finished is
easily calculated by realizing that the centroids $\vec{\hat X}$ as well as the coefficients
$J[m,x_rx_s]$ vanish for large values of $m=-2n$. Therefore, only terms with
$J[0,x_rx_s]$ that appear in Equation~\ref{eq:xxJ} survive in this limit. This leads to
\begin{eqnarray}\label{eq:Xasymp}
  \langle \hat x_1^2\rangle
  &=& \frac{1}{2}\left(J[0,x_1^2]+J[0,x_2^2]\right) = \eps_0+\frac{1}{2}(X_1^2+X_2^2)\nonumber\\
  \langle \hat x_1\hat x_2\rangle
  &=& 0\\
  \langle \hat x_2^2\rangle
  &=& \frac{1}{2}\left(J[0,x_1^2]+J[0,x_2^2]\right) = \eps_0+\frac{1}{2}(X_1^2+X_2^2)\ ,
      \nonumber 
\end{eqnarray}
where using Equation~\ref{eq:Jmrsr} for $m=0$ gives us $J[0,x_r,x_2]=(\eps_0\delta_{rs} +X_rX_s)$
and the asymptotic emittance $\hat\eps=\sqrt{\langle \hat x_1^2\rangle  \langle \hat x_2^2\rangle
-\langle \hat x_1\hat x_2\rangle^2}$. The asymptotic emittance growth then becomes $\hat\eps-\eps_0=(X_1^2+X_2^2)/2$
which is the Courant-Snyder invariant, written in coordinates of normalized phase space.
Expressed through physical coordinates, the centroid position $X$ and angle $X'$, the emittance
growth becomes
\begin{equation}
  \hat\eps-\eps_0= \frac{1}{2}\left(\gamma_x X^2 + 2\alpha_x X X' +\beta_x X^{\prime 2}\right)\ .
\end{equation}
This is not really
a surprise, because the amplitude-dependent tune shift does not change the oscillation
amplitudes of individual particles, such that the asymptotic emittance growth agrees with
the value caused by decoherence (Section~8.2 in~\cite{DECOH}) due to chromaticity and
momentum spread; only the transient behavior of the two processes differ.
\section{Mismatched beam}
\label{sec:mismatch}
%
\begin{figure}[tb]
\begin{center}
\includegraphics[width=0.97\textwidth]{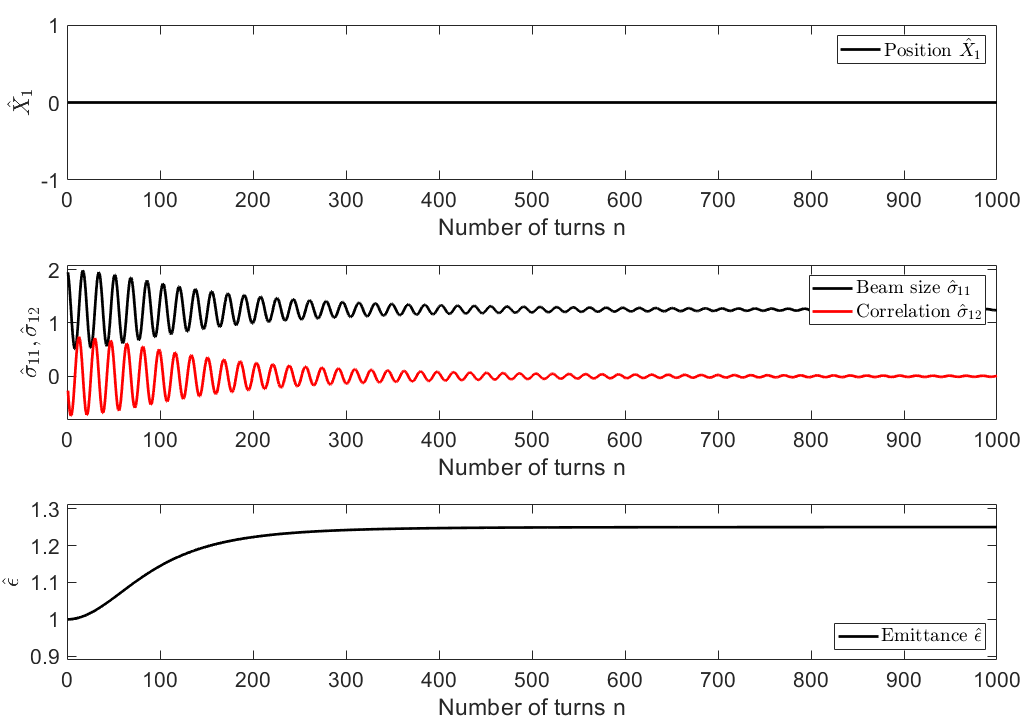}
\end{center}
\caption{\label{fig:emitm}The parameters $\hat X_1$, $\hat \sigma_{11}$, $\hat \sigma_{12}$, and emittance
  as a function of the turn number $n$ for a beam that is injected on axis, but with a beta function
  $\beta_0$ that is twice the matched value $\beta.$ All other parameters are equal to those
  used in Figure~\ref{fig:emit}. The vertical axes are normalized to appropriate powers of $\eps_0$.}
\end{figure}
In this section we explore the decoherence in one plane ($d=2$) of a mismatched beam that is injected on-axis
($\vec X= 0$) into the ring. In this case $\vec Y= 0$ and $\psi(m)=0$ from Equation~\ref{eq:psim},
which causes $J[m,x_rx_s]$ to simplify to
\begin{equation}
  J[m,x_rx_s]=\frac{1}{\sqrt{\det(\mathbf{1}+2im\kappa\sigma)}}
  \left(\mathbf{1}+2im\kappa\sigma\right)^{-1}\sigma\ .
\end{equation}
Moreover, we have $\vec{\hat X}= 0$. This makes calculating
the beam matrix $\hat\sigma$ and the emittance $\hat\eps$ straightforward. Figure~\ref{fig:emitm} shows
the result in position $\hat X_1$ (top), sigma matrix elements $\hat\sigma_{11}$ and $\hat\sigma_{12}$
(middle) and the emittance $\hat\eps$ (bottom) for an injected beam that has initial emittance
unity. We assume $\alpha=0$, but significantly increase the beta function to twice the value of the matched beam.
All other parameters are equal to those already used in Figure~\ref{fig:emit}. We see that the
beam size $\hat\sigma_{11}$ and correlation $\hat\sigma_{12}$ oscillate but this motion slowly
decoheres and reaches a new equilibrium value. At the same time, the emittance increases and also
settles towards a new, and larger, equilibrium value.
\par
\begin{figure}[tb]
  \begin{center}
    \includegraphics[width=0.97\textwidth]{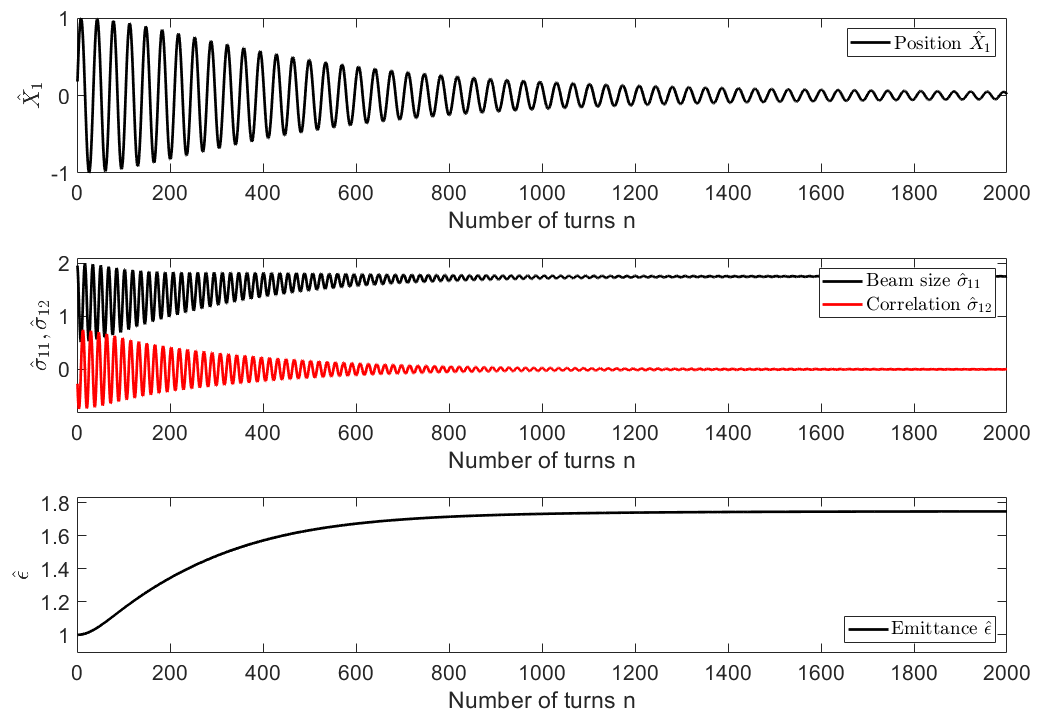}
  \end{center}
  \caption{\label{fig:emitms}The same parameters that are shown in Figure~\ref{fig:emitm},
    but with an additional steering error at injection $X_2=1.$ The slow decrease of $\hat X_1$
    motivates the extended range of turns.}
\end{figure}

Figure~\ref{fig:emitms} shows a simulation with parameters used in Figure~\ref{fig:emitm}, only the
initial value of $X_2$ is set to $X_2=1$. We see that $\hat X_1$ (top panel) performs betatron
oscillation with slowly decreasing amplitude, which motivates the increased range of turns shown.
Qualitatively, $\hat \sigma_{11}$ and $\hat \sigma_{12}$ (middle) show similar behavior to that
in Figure~\ref{fig:emitm}. Likewise, the emittance (bottom) increases to a new equilibrium value
that is, however, larger than the one on Figure~\ref{fig:emitm} due to the non-zero value of $X_2.$
\par
These new equilibrium values are easily calculated from Equations~\ref{eq:x1J} and~\ref{eq:xxJ}.
As before, realizing that all $J[m,x_rx_s]$ asymptotically vanish, this leaves us with
\begin{eqnarray}
  \langle \hat x_1^2\rangle
  &=& \frac{1}{2}\left(J[0,x_1^2]+J[0,x_2^2]\right)
      = \frac{1}{2}(\sigma_{11}+\sigma_{22})+\frac{1}{2}(X_1^2+X_2^2)\nonumber\\
  \langle \hat x_1\hat x_2\rangle
  &=& 0\\
  \langle \hat x_2^2\rangle
  &=& \frac{1}{2}\left(J[0,x_1^2]+J[0,x_2^2]\right)
      = \frac{1}{2}(\sigma_{11}+\sigma_{22})+\frac{1}{2}(X_1^2+X_2^2)\ ,
      \nonumber 
\end{eqnarray}
which is valid even for non-zero initial displacement $\vec X$. Here $(X_1^2+X_2^2)/2$ is again the Courant-Snyder
invariant of the centroid. Moreover, $\sigma_{jk}$ is the beam matrix of the injected beam in normalized
coordinates, which is related to the beam matrix in physical coordinates $\tilde\sigma$ by
\begin{equation}\label{eq:AsigA}
  \left(\begin{array}{cc}\sigma_{11} &\sigma_{12} \\ \sigma_{12} & \sigma_{22}\end{array}\right)
  ={\cal A}_x\tilde\sigma{\cal A}_x^{\top}
  \qquad\mathrm{with}\qquad
  \tilde\sigma=\eps_0\left(\begin{array}{cc}\beta_0 &-\alpha_0 \\ -\alpha_0 & \gamma_0\end{array}\right)\ ,
\end{equation}
where $\eps_0$ is the emittance and of the injected beam, $\alpha_0$, $\beta_0$, and $\gamma_0$
its Twiss parameters, and ${\cal A}_x$ is defined in Equation~\ref{eq:calA}. Evaluating
this expression and calculating $(\sigma_{11}+\sigma_{22})/2$ we arrive at
\begin{equation}\label{eq:Bmag}
  \frac{1}{2}(\sigma_{11}+\sigma_{22}) = \eps_0 B_{mag}
  \quad\mathrm{with}\quad
  B_{mag} = \frac{1}{2}\left[\left(\frac{\beta_0}{\beta_x}+\frac{\beta_x}{\beta_0}\right)
    +\beta_x\beta_0\left(\frac{\alpha_x}{\beta_x}-\frac{\alpha_0}{\beta_0}\right)^2\right]
\end{equation}
where we see that $B_{mag}$ is the factor by which the emittance of the injected beam is
asymptotically increased by decoherence after injecting a mismatched beam.
Summarily we find that the asymptotic emittance due to a displaced injected centroid and
mismatched beam matrix becomes
\begin{equation}\label{eq:epsasymp}
\hat\eps =\eps_0 B_{mag}+ \frac{1}{2}\left(\gamma_x X^2 + 2\alpha_x X X' +\beta_x X^{\prime 2}\right)
\end{equation}
with $B_{mag}$ defined in Equation~\ref{eq:Bmag} and the Twiss parameters of the ring $\alpha_x,$
$\beta_x,$ and $\gamma_x$. On-axis injection with the ratio of $\beta_0/\beta_x=2$ and
$\alpha=\alpha_0=0$, which is used in the simulation shown in Figure~\ref{fig:emitm}, leads to
$B_{mag}=1.25$, which agrees with the observed emittance growth visible on the bottom panel. Likewise,
additionally setting $X_2=1$ increases the emittance to $\hat\eps=B_{mag}\eps_0+X_2^2/2=1.75\,\eps_0$, which
agrees with the final value shown on the bottom panel in Figure~\ref{fig:emitms}.
%
\section{Transverse coupling}
\label{sec:coup}
%
For $d=4$, Equation~\ref{eq:sigmahat}, with $J[m,x_r,x_s]$ defined in Equation~\ref{eq:Jmrs},
describes the dynamics of a $4\times 4$ coupled beam matrix $\tilde\sigma$ that is injected
into a ring. In order to analyze it in a systematic way, we base our description on the
parameterization of coupled transfer matrices from~\cite{EDTENG,SAGRUB} and write
$\tilde\sigma$ as
\begin{equation}\label{eq:tilsig}
  \tilde\sigma=T^{-1}\tilde{\cal A}^{-1}\bar\eps\left(\tilde{\cal A}^{-1}\right)^{\top}\left(T^{-1}\right)^{\top}
  \quad\mathrm{with}\quad
\tilde{\cal A} = \left(\begin{array}{cc} \tilde{\cal A}_a & 0 \\ 0 & \tilde{\cal A}_b\end{array}\right)
\ \mathrm{and}\
\tilde{\cal A}_{a} = \left(\begin{array}{cc} \frac{1}{\sqrt{\beta_{a}}} & 0 \\
                       \frac{\alpha_{a}}{\sqrt{\beta_{a}}} & \sqrt{\beta_{a}}\end{array}\right),
\end{equation}
where $\tilde{\cal A}_{b}$ is defined analogously. Moreover, $\bar\eps=\diag(\eps_a,\eps_a,\eps_b,\eps_b)$
contains the emittances of two eigenmodes. $T$ and its inverse $T^{-1}$ describe transverse
coupling and are given by
 \begin{equation}\label{eq:tmatrix}
  T=\left( \begin{array}{cc} g\mathbf{1} & -C \\ C^+ & g\mathbf{1} \end{array} \right)
  \qquad \mathrm{and}\qquad
  T^{-1}=\left( \begin{array}{cc} g\mathbf{1} & C \\ -C^+ & g\mathbf{1} \end{array} \right) ,
\end{equation}
with the $2\times 2$ identity matrix $\mathbf{1}$, the $2\times 2$ coupling matrix $C$, its symplectic
conjugate $C^{+}=C^{-1}\det C$, and the scalar $g$, which satisfies $g^2=1-\det C$~\cite{SAGRUB}.
\par
We now transform the injected beam matrix $\tilde\sigma$, which is given in physical coordinates
to the coordinates of normalized phase space in the ring, which we call $\sigma$. Analogously to
what we did in Equation~\ref{eq:AsigA}, we transform it with ${\cal A}$, which has the same structure
as $\tilde{\cal A}$ from Equation~\ref{eq:tilsig}, but contains the Twiss parameters at the injection
point of the ring. We then obtain
\begin{equation}
  \sigma={\cal A}\tilde \sigma {\cal A}^{\top}
  = {\cal A} T^{-1} \tilde{\cal A}^{-1}\bar\eps\left(\tilde{\cal A}^{-1}\right)^{\top}\left(T^{-1}\right)^{\top} {\cal A}^{\top}
  ={\cal A} T^{-1} \tilde{\cal A}^{-1}\bar\eps \left({\cal A} T^{-1} \tilde{\cal A}^{-1}\right)^{\top}\ .
\end{equation}
Let us first calculate
\begin{equation}
  K={\cal A} T^{-1} \tilde{\cal A}^{-1}=
  \left(
    \begin{array}{cc}
      g{\cal A}_x\tilde{\cal A}_a^{-1} &{\cal A}_xC\tilde{\cal A}_b^{-1}\\
      -{\cal A}_yC^+\tilde{\cal A}_a^{-1} & g{\cal A}_y\tilde{\cal A}_b^{-1}
    \end{array}
  \right)
\end{equation}
which we use to calculate $\sigma=K\bar\eps K^{\top}$ and find the top-left $2\times 2$ submatrix
of $\sigma$ to be
\begin{equation}\label{eq:scoup}
 \left(\begin{array}{cc} \sigma_{11} & \sigma_{12}\\\sigma_{12} &\sigma_{22}\end{array}\right)
 =g^2\eps_a{\cal A}_x\tilde{\cal A}_a^{-1}\left({\cal A}_x\tilde{\cal A}_a^{-1}\right)^{\top}
 +\eps_b{\cal A}_xC\tilde{\cal A}_b^{-1} \left({\cal A}_xC\tilde{\cal A}_b^{-1}\right)^{\top}
\end{equation}
from which we calculate the asymptotically achievable emittance with $(\sigma_{11}+\sigma_{22})/2$,
just as we did in the previous section. The lower-right submatrix contains a similar expression that
describes the vertical plane from which we can calculate the asymptotically achievable vertical
emittance $(\sigma_{33}+\sigma_{44})/2$.
\par
\begin{figure}[tb]
\begin{center}
\includegraphics[width=0.97\textwidth]{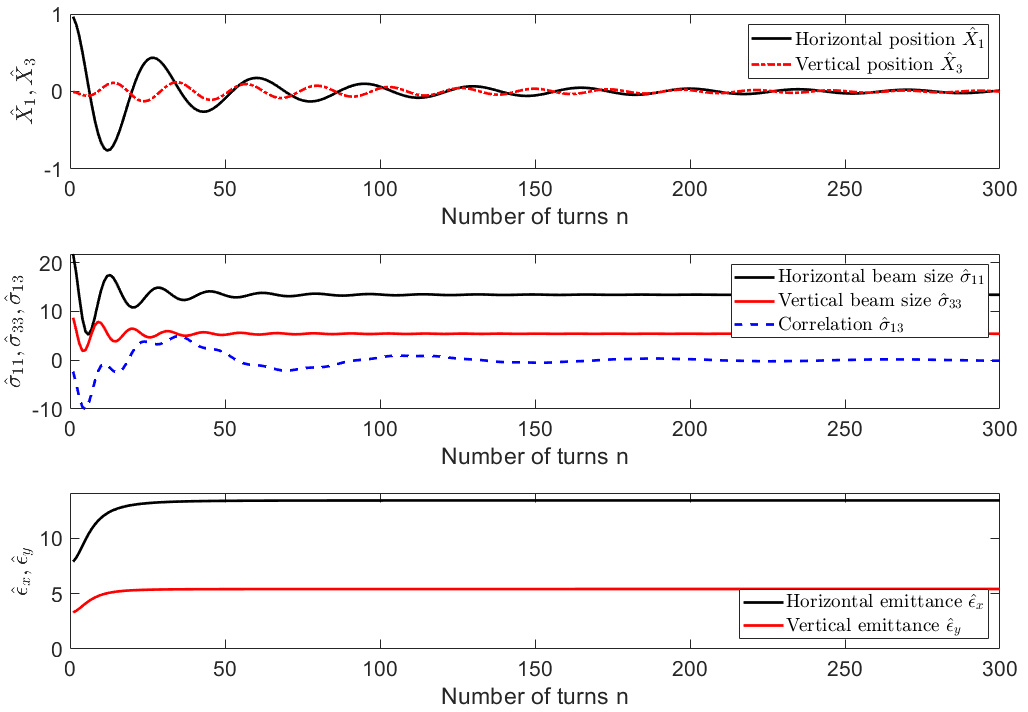}
\end{center}
\caption{\label{fig:coup}The horizontal and vertical beam positions $\hat X_1$ and $\hat X_3$ (top), beam matrix elements
  (middle) and emittance (bottom) as a function of the number of turns for a beam with
  initial emittance ratio $\eps_a/\eps_b=10$, initial beta mismatch, and displacement. The
  beam is rotated by $\eta=30^o$. The initial mismatch decoheres and the emittance reaches
  its asymptotic value, given by Equation~\ref{eq:ec2}. }
\end{figure}
We now consider the special case where $C$ stems from a coordinate rotation with angle~$\eta$.
This leads to $g=\cos\eta$ and $C=-\mathbf{1}\sin\eta$. Inserting $g$ and $C$ into
Equation~\ref{eq:scoup}, we obtain
\begin{equation}\label{eq:scoupx}
 \left(\begin{array}{cc} \sigma_{11} & \sigma_{12}\\\sigma_{12} &\sigma_{22}\end{array}\right)
 =\eps_a{\cal A}_x\tilde{\cal A}_a^{-1} \left({\cal A}_x\tilde{\cal A}_a^{-1}\right)^{\top}\cos^2(\eta)
 +\eps_b{\cal A}_x\tilde{\cal A}_b^{-1} \left({\cal A}_x\tilde{\cal A}_b^{-1}\right)^{\top}\sin^2(\eta)\ .
\end{equation}
The combination of matrices in the second term evaluates to
\begin{equation}\label{eq:scoupxx}
{\cal A}_x\tilde{\cal A}_b^{-1} \left({\cal A}_x\tilde{\cal A}_b^{-1}\right)^{\top}  
= \left(
  \begin{array}{cc}
    \frac{\beta_b}{\beta_x} & \frac{\alpha_x\beta_b}{\beta_x} -\alpha_b \\
    \frac{\alpha_x\beta_b}{\beta_x} -\alpha_b
                                  & \frac{\alpha_x^2\beta_b}{\beta_x}
                                    - 2\alpha_x\alpha_b+\frac{1+\alpha_b^2}{\beta_b}\beta_x
  \end{array}
\right)          
\end{equation}
and to a similar expression for the first term after replacing $\beta_b$ and $\alpha_b$ by
$\beta_a$ and $\alpha_a$, respectively. From the sum of the diagonal elements, we
obtain for the asymptotically achievable emittances in the horizontal and the vertical plane
\begin{eqnarray}\label{eq:emitcoup}
  \frac{1}{2}(\sigma_{11}+\sigma_{22})&=&\eps_a\cos^2(\eta)B_{mag}(\beta_x,\beta_a) +\eps_b\sin^2(\eta)B_{mag}(\beta_x,\beta_b)\nonumber\\
  \frac{1}{2}(\sigma_{33}+\sigma_{44})&=&\eps_b\cos^2(\eta)B_{mag}(\beta_y,\beta_b) +\eps_a\sin^2(\eta)B_{mag}(\beta_y,\beta_a)\\
  \quad\mathrm{with}\quad
    B_{mag}(\beta_x,\beta_b) &=& \frac{1}{2}\left[\left(\frac{\beta_x}{\beta_b}+\frac{\beta_b}{\beta_x}\right)
    +\beta_x\beta_b\left(\frac{\alpha_x}{\beta_x}-\frac{\alpha_b}{\beta_b}\right)^2\right]\ ,\nonumber
\end{eqnarray}
where we do not write out the dependence on $\alpha_x$ and $\alpha_b$ in the definition
of $B_{mag}$, whose definition from Equation~\ref{eq:Bmag} is repeated here for convenience.
In Equation~\ref{eq:emitcoup} it contains different combinations of horizontal and vertical
Twiss parameters of the injected beam and those at the point of
injection into the ring. It describes the influence of the Twiss parameters on the decoherence,
which is smallest ($B_{mag}=1$), if the Twiss parameters in the horizontal and vertical plane
of the injection line and the ring are equal. Summarily, the asymptotic emittance growth,
including the effect of initial displacement, in the horizontal plane then turns out to be
\begin{equation}\label{eq:ec2}
  \hat\eps_x=\eps_aB_{mag}(\beta_x,\beta_a)\cos^2(\eta) +\eps_bB_{mag}(\beta_x,\beta_b)\sin^2(\eta)
  +\frac{1}{2}\left(X_1^2+X_2^2\right)
\end{equation}
and a corresponding equation for the vertical emittance.
\par
Figure~\ref{fig:coup} shows the turn-by-turn evolution of a beam with initial emittance ratio
of $\eps_a/\eps_b=10$ that is coupled by a coordinate rotation with $\eta=30^o$. The Twiss
parameters of the injected beam are $\beta_a=\beta_b=3\,$m, and $\alpha_a=\alpha_b=0$, which
makes $B_{mag}(\beta_x,\beta_a)=B_{mag}(\beta_x,\beta_b)=5/3$.
Moreover, the beam is injected with an initial offset $X_1=1$. The tunes are $0.028$ in the
horizontal and $0.041$ in the vertical plane and the detuning parameters are $\kappa_{xx}=10^{-3}$,
$\kappa_{yy}=2\times 10^{-3}$ and $\kappa_{xy}=5\times 10^{-4}$. We observe in the upper panel that
the beam initially performs horizontal betatron oscillations with decreasing amplitude, but the
coupled beam matrix also causes the vertical centroid $\hat X_3$ to oscillate. Likewise, the
horizontal and vertical beam sizes, both shown in the middle panel, initially oscillate, but
rapidly decohere, before settling on their equilibrium value. The correlation $\hat\sigma_{13}$,
derived in Appendix~\ref{sec:appA},
shows a more complicated pattern, because it oscillates with sum and difference frequency of
the horizontal and vertical tune before also reaching its equilibrium value zero.
The bottom panel shows the
horizontal and vertical emittances increasing from their initial value, which is given by the
projected emittance of the coupled beam at injection. Decoherence causes the emittances to
asymptotically reach $\hat\eps_x=13.4$ and $\hat\eps_y=5.41$, consistent with the values
calculated from Equation~\ref{eq:ec2}.
\section{Dispersion}
\label{sec:disp}
In this section we consider the asymptotic emittance growth due to a mismatched and potentially
coupled dispersion with $d=4$. Here we treat dispersion errors $\vec D$ as a momentum-dependent offset of
the centroid, such that we just replace $\hat X$ by $\vec D\delta$ in Equation~\ref{eq:Xasymp}.
Subsequently averaging over $\delta$ gives us the emittance growth as
\begin{equation}
  \Delta\hat\eps=\frac{1}{2}\left( D_1^2+ D_2^2\right)\sigma_{\delta}^2\ ,
\end{equation}
where $\sigma_{\delta}$ is the relative momentum spread in the ring. The dispersion
errors $ \vec D$ in normalized phase space are given by
\begin{equation}
   \vec D=( D_1, D_2, D_3, D_4)^{\top}
  ={\cal A} T \left(\begin{array}{c} \vec D_x\\ \vec D_y \end{array}\right)
  = \left(\begin{array}{cc}g{\cal A}_x \vec D_x -{\cal A}_x C \vec D_y\\
     {\cal A}_y C^+ \vec D_x +  g{\cal A}_y \vec D_y \end{array}\right)\ ,    
\end{equation}
where ${\cal A}$ from Equation~\ref{eq:tilsig} contains the Twiss parameters and $T$ from
Equation~\ref{eq:tmatrix} describes transverse coupling. These two matrices transform the physical
dispersions $\vec D_x=( D_x, D'_x)^{\top}$ and $ \vec D_y=( D_y, D'_y)^{\top}$
in the horizontal and vertical
plane of the transfer line into the normalized phase space of the ring. Evaluating $ D_1^2+ D_2^2$
then leads to
\begin{equation}\label{eq:dispa}
   D_1^2+ D_2^2 = g^2 \vec  D_x^{\top}{\cal A}_x^{\top}{\cal A}_x \vec D_x
  - 2 g \vec D_y^{\top}C^{\top}{\cal A}_x^{\top}{\cal A}_x \vec D_x
  + \vec D_y^{\top}C^{\top}{\cal A}_x^{\top}{\cal A}_xC\vec D_y
\end{equation}
and a similar expression for $ D_3^2+ D_4^2$ that describes the emittance growth in the vertical plane.
Equation~\ref{eq:dispa} is valid for any coupling matrix $C$, but if we specifically evaluate it
for a coordinate rotation with $g=\cos\eta$ and $C=-\mathbf{1}\sin\eta$, we find
\begin{equation}\label{eq:dispb}
  D_1^2+D_2^2=\cos^2(\eta){\cal H}_x(\vec D_x,\vec D_x) + 2\sin(\eta)\cos(\eta){\cal H}_x(\vec D_y,\vec D_x)
                 +\sin^2(\eta){\cal H}_x(\vec D_y,\vec D_y)
\end{equation}                 
where
\begin{equation}
  {\cal H}_x(\vec D_y,\vec D_x)=\gamma_xD_xD_y+\alpha_x(D_yD'_x+D'_yD_x)+\beta_xD'_xD'_y
\end{equation}
is the generalization of the quantity ${\cal H}_x$ that appears in the fifth radiation
integral~\cite{HELM,ASVZ}.
\par
For $\eta=0$, Equation~\ref{eq:dispb} characterizes the emittance growth due to a dispersion error
$\vec D_x$ in the horizontal plane. The emittance growth then turns out to be
\begin{equation}
  \Delta\hat\eps = \frac{1}{2}{\cal H}_x(\vec D_x,\vec D_x)\sigma_{\delta}^2
  =\frac{1}{2}\left(\gamma_xD_x^2+2\alpha_xD_xD'_x+\beta_xD^{\prime 2}_x\right)\sigma_{\delta}^2\ ,
\end{equation}
which agrees with the expression derived in~\cite{ARDUINI}.
\section{Chromaticity}
\label{sec:chroma}
The decoherence of an unbunched beam with rms momentum spread $\sigma_{\delta}$ and a finite
chromaticity $Q'$  can be included in our framework by adding $\mu'_x\delta=2\pi Q'_x\delta$
to the phase advance per turn $\phi_x$ from Equation~\ref{eq:adps}. This gives us
\begin{equation}\label{eq:phidelta}
  \phi_x= \mu_x  + \vec x^{\top}\bar\kappa_x\vec x+ \mu'_x\delta
  \quad\mathrm{with}\quad
  \psi(\delta)=\frac{1}{\sqrt{2\pi}\sigma_{\delta}}e^{-\delta^2/2\sigma^2_{\delta}}\ .
\end{equation}
Instead of just averaging over the transverse phase-space coordinates in Equation~\ref{eq:INP},
we now also have to average over the momentum $\delta$ with distribution $\psi(\delta)$.
The integral factorizes into one part that depends on $x_1$ and $x_2$ and a second,
momentum-dependent part $D(n)$, given by
\begin{equation}\label{eq:chrff}
  D(n)=\int e^{-i n \mu'_x\delta } e^{-\delta^2/2\sigma^2_{\delta}} d\delta = e^{-\mu^{\prime 2}\sigma^2_{\delta}n^2/2}\ ,
\end{equation}
which multiplies all integrals $I[n,\vec p]$.
\par
For bunched beams that perform synchrotron oscillations with frequency $\nu_s$,
the betatron phase advance after $n$ turns is given by~\cite{Meller,MINTY}
\begin{equation}
  n\phi_x= n\mu_x  + n\vec x^{\top}\bar\kappa_x\vec x+ \zeta(n)
  \quad\mathrm{with}\quad
  \zeta(n)=\frac{\mu'_x\delta}{\pi\nu_s}\sin(\pi\nu_sn)\cos(\pi\nu_sn+\eta_0)\ ,
\end{equation}
where $\eta_0$ is the initial phase of the synchrotron oscillations. Averaging over
$\eta_0$ and $\delta$ with the momentum distribution from Equation~\ref{eq:phidelta}
results in the form factor~\cite{Meller,MINTY}
\begin{equation}\label{eq:chrffb}
  D(n)=\exp\left[-2\left(\frac{\mu'\sigma_{\delta}}{2\pi\nu_s}\right)^2\sin^2(\pi\nu_sn)\right]\ .
\end{equation}
The form factor $D(n)$, either from Equation~\ref{eq:chrff} for unbunched beams or from
Equation~\ref{eq:chrffb} for bunched beams, becomes a multiplicative factor for $I[n,\vec p]$
that carries through all the way to Equation~\ref{eq:evoX}, where it modulates the right-hand
side. In the same fashion, all $J[-2n,x_rx_2]$ in Equations~\ref{eq:x1J},~\ref{eq:xxJ},
and~\ref{eq:xxxJ} assume an additional factor $D(n)^4$, because the step from $n$ to $2n$
doubles $\zeta(n)$, which is equivalent to doubling $\mu'_x$ that causes the exponent
of $D(n)$ to quadruple. Apart from these additional factors, all other equations remain
unchanged. In particular, the asymptotic equilibrium values of the beam matrix and the
emittance, which are multiplied by powers of $D(0)=1$, from Equations~\ref{eq:epsasymp}
and~\ref{eq:emitcoup} remain unaffected. Only the temporal evolution towards the equilibrium
is modulated by the powers of $D(n)$ which prepend the $J(\pm 2n,x_rx_s)$.
\section{Tolerances}
%
\begin{table}[tb]
  \caption{\label{tab:sps}The tolerance levels for mismatch and steering errors
    for the injection into the SPS.
    The nominal emittance is $\eps_0=1.26\times 10^{-7}\,$m\,rad and the Twiss
    parameters at the injection point are $\beta=44.5\,$m and $\alpha=-0.96$.
  }
  \begin{center}
    \begin{tabular}{c|cccc}
      Tolerance level  & $\Delta\beta/\beta$ & $\Delta\alpha$ & $\Delta X$ [mm] & $\Delta X'$ [$\mu$rad] \\
      \hline
      1\,\%   & 0.14 & 0.14 & 0.24 & 7.5 \\
      5\,\%   & 0.32 & 0.32 & 0.54 & 16.8      
    \end{tabular}
  \end{center}
\end{table}
Here we analyze the requirements for the steering errors and the Twiss parameters of
an injected beam to cause an emittance growth of less than 1\,\% and 5\,\%. To do so,
we expand Equation~\ref{eq:epsasymp} up to second order in the deviations from their
respective design values $\Delta\beta=\beta_0-\beta$, $\Delta\alpha=\alpha_0-\alpha$,
$\Delta X$, and $\Delta X'$ and find for the asymptotic emittance increase 
\begin{equation}
  \hat\eps-\eps_0=\frac{1}{2}\left(\frac{\Delta\beta}{\beta}\right)^2 +\frac{1}{2}\Delta\alpha^2
  +\frac{\gamma}{2} \Delta X^2 + \frac{\beta}{2} \Delta X^{\prime\,2}
\end{equation}
with $\gamma=(1+\alpha^2)/\beta$. As example, we use the horizontal injection from the
TT10 transfer line into the
SPS~\cite{SPS} when it serves beams to the LHC. In this configuration the horizontal
Twiss parameters~\cite{SPSTWISS} at the injection point are $\beta=44.5\,$m and $\alpha=-0.96$.
Moreover, the emittance is
$\eps_0=1.26\times 10^{-7}\,$m\,rad. The tolerance levels that increase the asymptotic
emittance by 1\,\% and 5\,\% are shown in Table~\ref{tab:sps}. We find that the error
tolerances for the Twiss parameters are fairly relaxed; even errors of $\Delta\beta/\beta$
or $\Delta\alpha$ in the 10\,\% range increase the emittance by less than 1\,\%. On the
other hand, owing to the relatively large value of $\beta$ at the injection point, 
steering errors $\Delta X'$ exceeding $20\,\mu$rad lead  to increased emittances above
the 5\,\% level.
\section{Conclusion}
We derived evolution equations for the first and second moments of an coupled arbitrary Gaussian
phase-space distribution that initially is mismatched, displaced, and has mismatched dispersion
under the influence of
decoherence due to amplitude-dependent tune shift. The well-known results from~\cite{Meller}
and~\cite{MINTY} for the amplitude dependence of the first and seconds moments after an
initial displacement of a matched beam are reproduced. Our results go beyond~\cite{Meller}
and~\cite{MINTY}, because the initial beam can have an arbitrary Gaussian distribution,
which includes transverse coupling, and does not need to be matched. We then calculate the
temporal evolution of the second moments,
the beam sizes, and the emittance. Moreover, we calculate the emittance in the asymptotic
limit and find it to agree with the emittance growth due to chromatic effects. Finally we
analyzed tolerances for the injection and used the SPS as an illustration.
\section*{Acknowledgements}
We acknowledge financial support from Uppsala University (VZ) and from CERN (EW).
We gratefully acknowledge discussions with Francesco Velotti and the hospitality of the ABT group
at CERN.
%
\bibliographystyle{plain}

\begin{thebibliography}{M}
%
\bibitem{Meller}
  R. Meller, A. Chao, J. Peterson, S. Peggs, M. Furman, {\em Decoherence of kicked beams,} SSC-N-360, May 1987.
\bibitem{SYLEE}
  S.Y. Lee, {\em Decoherence of the kicked beams II,} SSC-N-749, February 1991.
\bibitem{Sargsyan}
  A. Sargsyan, {\em Transverse decoherence of the kicked beams due to amplitude and
    chromaticity tune shifts,} Nucl. Inst. Meth. A638 (2011) 15.
\bibitem{MINTY}
  M. Minty, A. Chao, W. Spence, {\em Emittance growth due to decoherence and wakefields,}
  Proceedings of the 1995 Particle Accelerator Conference in Dallas (1995) 3037.
\bibitem{HANDBOOK}
  M. Furman, {\em Decoherence,} Section 2.3.10 in A. Chao, M. Tigner, {\em Handbook of
    Accelerator Physics and Engineering, 1st ed.}, World Scientific, Singapore,  1999.
\bibitem{VERDIER}
  A. Verdier, {\em Cross anharmonicity,} AIP Conference Proceedings 344 (1995) 269;
  doi: https://doi.org/10.1063/1.48995.
\bibitem{ZJ}
  Section 1.1 in J. Zinn-Justin, {\em Quantum field theory and critical phenomena, 4th ed.},
  Clarendon Press, Oxford, 2002.
\bibitem{MATLAB}
  MATLAB web page:  \url{https://www.mathworks.com}
\bibitem{GITHUB}
  Github repository with the software for the simulations: \url{https://github.com/volkziem/InjectionDecoherence}.
\bibitem{DECOH}
  V. Ziemann, {\em Hands-On Accelerator Physics Using MATLAB,} CRC Press, Boca Raton, 2019.
\bibitem{EDTENG}
  D. Edward and L.Teng, {\em Parametrization of linear coupled motion in periodic
  systems,} IEEE Trans.Nucl.Sci. 20, 885 (1973).
\bibitem{SAGRUB}
  D. Sagan, D. Rubin, {\em Linear Analysis of coupled lattices,} Physical Review Special
  Topics--Accelerators and Beams 2 (1999) 074001.
\bibitem{HELM}
  R. Helm, M. Lee, P. Morton, {\em Evaluation of synchrotron radiation integrals,} IEEE
  Trans.Nucl.Sci. 20 (1973) 900. 
\bibitem{ASVZ}
  V. Ziemann, A. Streun, {\em Equilibrium parameters in coupled storage ring lattices and practical
    applications,} Physical Review Accelerators and Beams, accepted; see also arXiv:2201.11025.
\bibitem{ARDUINI}
  G. Arduini, P. Raimondi, {\em Transverse emittance blow-up due to injection errors,}
  CERN SL-Note-99-022, 1999.
\bibitem{SPS}
  P. Collier et al., {\em SPS as injector for LHC: Conceptual design report,}
  CERN SL-97-07 DI, CERN, 1997.
\bibitem{SPSTWISS}
  CERN Optics Repository, Q20 optics function for SPS injection, available at
  \url{https://acc-models.web.cern.ch/acc-models/tls/2021/sps_injection/tt2tt10_lhc_q20/stitched/}
  [retrieved 2022/03/15]
\end{thebibliography}

\appendix
\section{Second moments}
\label{sec:appA}
In Equation~\ref{eq:x1sq}, we only show one of the second-order moments. The other
two that are needed for the horizontal plane are calculated in a similar fashion from
\begin{eqnarray}
  \langle \hat x_1\hat x_2\rangle 
  &=& \left\langle\left(x_1\cos n\phi_x+x_2\sin n\phi_x\right)
      \left(-x_1\sin n\phi+x_2\cos n\phi\right)\right\rangle\nonumber\\
  \langle \hat x_2^2\rangle 
  &=& \left\langle\left(-x_1\sin n\phi_x+x_2\cos n\phi_x\right)^2\right\rangle\
\end{eqnarray}
where the angle brackets denote averaging over the Gaussian from Equation~\ref{eq:gauss}
in $d$ dimensions. Following steps similar to those leading to Equation~\ref{eq:x1J} brings us to 
\begin{eqnarray}\label{eq:xxJ}
  \langle \hat x_1^2\rangle
  &=& \frac{1}{4}\left( 2J[0,x_1^2]+J[-2n,x_1^2]+J[2n,x_1^2] \right)\nonumber\\
  &&\quad    -\frac{i}{2}\left( J[-2n,x_1x_2]  - J[2n,x_1x_2]\right)\nonumber\\
  &&\quad +\frac{1}{4}\left(2J[0,x_2^2]-J[-2n,x_2^2]-J[2n,x_2^2]\right)\nonumber\\
  \langle \hat x_1\hat x_2\rangle
  &=&-\frac{1}{4i}\left(J[-2n,x_1^2]-J[2n,x_1^2]\right)\nonumber\\
  &&\quad  +\frac{1}{2}\left(J[-2n,x_1x_2]+J[2n,x_1x_2]\right)\\
  &&\quad   +\frac{1}{4i}\left(J[-2n,x_2^2]-J[2n,x_2^2]\right)\nonumber\\
  \langle \hat x_2^2\rangle
  &=& \frac{1}{4}\left(2J[0,x_1^2]-J[-2n,x_1^2]-J[2n,x_1^2]\right)\nonumber\\
  &&\quad +\frac{i}{2}\left(J[-2n,x_1x_2]-J[2n,x_1x_2]\right)\nonumber\\
  &&\quad+\frac{1}{4}\left(2J[0,x_2^2]+J[-2n,x_2^2]+J[2n,x_2^2]\right)\ ,\nonumber
\end{eqnarray}
where, for completeness, we also show the expression for $\langle \hat x_1^2\rangle$ 
from Equation~\ref{eq:x1J}. We can simplify these expressions further by noting that
\begin{equation}
  J[-m,p]+J[m,p]=\langle e^{im\phi_x}p\rangle+\langle e^{-im\phi_x}p\rangle
  =2\Real \langle e^{im\phi_x}p\rangle=2 \Real(J[-m,p])
\end{equation}
and likewise
\begin{equation}
  J[-m,p]-J[m,p]=2i\Imag(J[-m,p])\ ,
\end{equation}
which allows us to write
\begin{eqnarray}\label{eq:xxxJ}
  \langle \hat x_1^2\rangle
  &=&\frac{1}{2}\left(J[0,x_1^2]+\Real(J[-2n,x_1^2])\right) +\Imag(J[-2n,x_1x_2])\nonumber\\
  &&\quad +\frac{1}{2}\left(J[0,x_2^2]-\Real(J[-2n,x_2^2])\right)\nonumber\\
  \langle \hat x_1\hat x_2\rangle
  &=& -\frac{1}{2}\Imag(J[-2n,x_1^2])+\Real(J[-2n,x_1x_2])+\frac{1}{2}\Imag(J[-2n,x_2^2])\nonumber\\
  \langle \hat x_2^2\rangle 
  &=&\frac{1}{2}\left(J[0,x_1^2]-\Real(J[-2n,x_1^2]\right) - \Imag(J[-2n,x_1x_2])\\
  &&\quad+\frac{1}{2}\left(J[0,x_2^2]+\Real(J[-2n,x_2^2]\right)\ .
\nonumber
\end{eqnarray}
\par
The second moments of the type $\langle\hat x_1\hat x_3\rangle$ arise if we consider
coupled motion and need special attention, because $\hat x_1$ oscillates with $\mu_x$
and $\hat x_3$ with $\mu_y$. Likewise the amplitude-dependent tuneshift in the
horizontal plane is given by $\vec x^{\top}\bar\kappa_x \vec x$ and by
$\vec x^{\top}\bar\kappa_y \vec x$ with $\bar\kappa_y=\diag(\kappa_{xy},\kappa_{xy},\kappa_{yy},\kappa_{yy})$
in the vertical plane. Since we will encounter
$J[m,p;\mu_x,\bk_x]$ from Equation~\ref{eq:Jmp} for different arguments $\mu_x$
and $\bk_x$, we specify all arguments henceforth when we calculate
$\langle\hat x_1\hat x_3\rangle$ for which we find
\begin{eqnarray}
  \langle\hat x_1\hat x_3\rangle
  &=& \langle \left[x_1\cos(n\phi_x) +x_2\sin(n\phi_x)\right]\left[x_3\cos(n\phi_y) +x_4\sin(n\phi_y)\right]\rangle
      \nonumber\\
  &=& \frac{1}{4}\langle x_1x_3\left[e^{in(\phi_x+\phi_y)}+e^{in(\phi_x-\phi_y)}+e^{-in(\phi_x-\phi_y)}+e^{-in(\phi_x+\phi_y)}\right]\rangle
      \nonumber\\
  && + \frac{1}{4i}\langle x_1x_4\left[e^{in(\phi_x+\phi_y)}-e^{in(\phi_x-\phi_y)}+e^{-in(\phi_x-\phi_y)}-e^{-in(\phi_x+\phi_y)}\right]\rangle
      \nonumber\\
  && + \frac{1}{4i}\langle x_2x_3\left[e^{in(\phi_x+\phi_y)}+e^{in(\phi_x-\phi_y)}-e^{-in(\phi_x-\phi_y)}-e^{-in(\phi_x+\phi_y)}\right]\rangle
      \nonumber\\
  && - \frac{1}{4}\langle x_2x_4\left[e^{in(\phi_x+\phi_y)}-e^{in(\phi_x-\phi_y)}-e^{-in(\phi_x-\phi_y)}+e^{-in(\phi_x+\phi_y)}\right]\rangle\\
  &=&\frac{1}{4}\left( 2\Real(J[-n,x_1x_3;\mu_x+\mu_y,\bk_x+\bk_y])+2\Real(J[-n,x_1x_3;\mu_x-\mu_y,\bk_x-\bk_y])\right)
      \nonumber\\
  &&+\frac{1}{4i}\left(2i\Imag(J[-n,x_1x_4;\mu_x+\mu_y,\bk_x+\bk_y]-2i\Imag(J[-n,x_1x_4;\mu_x-\mu_y,\bk_x-\bk_y]\right)
     \nonumber\\
  &&+\frac{1}{4i}\left(2i\Imag(J[-n,x_2x_3;\mu_x+\mu_y,\bk_x+\bk_y]+2i\Imag(J[-n,x_2x_3;\mu_x-\mu_y,\bk_x-\bk_y]\right)
     \nonumber\\
  &&-\frac{1}{4}\left(2\Real(J[-n,x_2x_4;\mu_x+\mu_y,\bk_x+\bk_y]-2\Real(J[-n,x_2x_4;\mu_x-\mu_y,\bk_x-\bk_y]\right)\ .
     \nonumber
\end{eqnarray}
The last equality is a sum of terms very much like those from Equation~\ref{eq:x1J}. Only here the phase advance
$\mu_x$ is replaced by $\mu_x\pm\mu_y$ and $\bk_x$ by $\bk_x\pm\bk_y$. We can therefore use the same MATLAB function
for $J[m,p;\mu_x,\bk_x]$ to work out $\langle\hat x_1\hat x_3\rangle$ and determine $\hat\sigma_{13}=
\langle\hat x_1\hat x_3\rangle-\hat X_1\hat X_3$ shown on the middle panel in Figure~\ref{fig:coup}.
\end{document}